\documentclass{iau}

\usepackage{amsmath}
\usepackage{graphicx}
\usepackage{multirow}

\begin{document}

\lefttitle{Gonz\'alez-Tor\`a et al.}
\righttitle{Can Wolf-Rayet stars be the missing ingredient to explain high-z He II ionizing radiation?}

\jnlPage{1}{7}
\jnlDoiYr{2021}
\doival{10.1017/xxxxx}

\aopheadtitle{Proceedings IAU Symposium}
\editors{A. Wofford,  N. St-Louis, M. Garcia \&  S. Simón-Díaz, eds.}

\title{Can Wolf-Rayet stars be the missing ingredient to explain high-z He II ionizing radiation?}

\author{G. Gonz\'alez-Tor\`a$^1$, A. A. C. Sander$^{1,2}$, E. Egorova$^1$, O. Egorov$^1$, M. Bernini-Peron$^1$, J. Josiek$^1$,  K. Kreckel$^1$, R. R. Lefever$^1$, V. Ramachandran$^1$, E. C. Schösser$^1$}
\affiliation{1) Zentrum für Astronomie der Universität Heidelberg, Astronomisches Rechen-Institut, Mönchhofstr. 12-14, 69120 Heidelberg}
\affiliation{2) Universität Heidelberg, Interdiszipliäres Zentrum für Wissenschaftliches Rechnen, 69120 Heidelberg, Germany}

\begin{abstract}
Classical Wolf-Rayet (WR) stars are hot, massive stars with depleted hydrogen. At low metallicities (Z), WN3-type WR stars have relatively thin winds and are major sources of ionizing flux. The detection of high-ionization emission lines in high-redshift ($z$) galaxies as well as nearby low-Z dwarf galaxies raises questions about the origin of He II ionizing radiation and its role in galaxy evolution, as stellar population models fail to reproduce the required fluxes. Low-Z WN3 stars may provide the missing contribution but are easily hidden in integrated light. Using the Local Volume Mapper, we compare resolved optical spectra of SMC WN3 stars with integrated regions, focusing on the broad He II $\lambda4686\,\AA$ line. We find stellar emission diluted within nebular regions, becoming undetectable when integrating over areas larger than 24 pc. Nonetheless, these stars emit enough ionizing photons to explain observed He II nebular emission, being strong candidates for the He II ionizing sources in low-Z and high-$z$ galaxies.

\end{abstract}


\maketitle

\section{Introduction}

\noindent
Nebular He II recombination emission lines are present in both high-$z$ \citep[e.g.,][]{Topping2025} and nearby low-metallicity ($Z$) galaxies \citep[e.g.,][]{Kehrig2018,Wofford2021}, implying the existence of sources that can produce a significant fraction of high energy photons ($\lambda<228\,$\AA) to ionize He$^{+}$ in their surrounding interstellar medium. 
Wolf-Rayet stars are hot ($T_{\mathrm{eff}}>30$\,kK) massive stars with very dense winds that present broad emission lines in their spectrum. Among them, a subclass of WR stars showing strong helium and nitrogen emission lines are defined as WN stars. Particularly, WN3 stars are WN stars showing emission from high ionization species (e.g., He II, N V), with many of them showing relatively thin winds among the WR regime. Spectral modeling has revealed them as important sources of He$^{+}$ ionizing flux \citep[e.g.,][]{Hainich2015,Sander2022}, which makes them particularly interesting as the WR subtype occurrence shifts to earlier types at lower $Z$ \citep[e.g.,][]{Crowther2007}.

While resolving individual stars in high-$z$ galaxies is unfeasible, we have a nearby low-$Z$ environment with a resolvable WN3 stellar population, the Small Magellanic Cloud \citep[SMC, $Z \approx 0.2\,Z_\odot$,][]{Bouret2003,Ramachandran2019}. Hosting 12 known, resolvable WR stars, we compare the intrinsic stellar properties and spectra of 6 WN3 in the SMC (AB\,1, 7, 9, 10, 11 and 12) with unresolved integral field unit (IFU) spectra from the SDSS V Local Volume Mapper survey \citep[LVM,][]{Drory2024}, focusing on the 4686 $\AA$ He II line, the strongest optical emission line for WN stars \citep[see also][]{Gonzalez-Tora2025}.

\section{Atmosphere analyses and IFU observations}

\noindent%
All the individual WN3 targets have optical (3700 - 6830 $\AA$) data from \citet{Foellmi2003}, and have been previously analyzed with the PoWR atmosphere code \citep{Graefener2002,Sander2015} in \citet{Hainich2015} or \citet{Shenar2016}. Despite having a WR classification, only the models for the WN3 star in AB\,7 filter yields a wind optical depth larger than unity, namely $\tau_F(R_\text{sonic}) = 2.1$. All single WN3 in the SMC analyzed in \citet{Hainich2015} show $\tau_F(R_\text{sonic}) < 1$, spanning between $0.29$ (AB\,1) and $0.8$ (AB\,10). This means that the winds of the single SMC WN3 stars are not globally optically thick. The transition in optical thickness further does not align with transparency in He\,II ionizing flux, as outlined by AB\,7 and the theoretical study by \citet{Sander2023}.
From the PoWR atmosphere models \citep[taken from][]{Hainich2015,Shenar2016}, we can compute the emerging ionizing fluxes Q$_{\mathrm{edge}}$ for an ion with a certain ionization edge frequency $\nu_{\mathrm{edge}}$ via 
\begin{equation}
    Q_{\mathrm{edge}}:= 4 \pi R_\ast^2 \int_{\nu_{\mathrm{edge}}}^{\infty} \frac{F_{\nu}}{h\nu} \,\mathrm{d}\nu,
\end{equation}
with $F_{\nu}$ denoting the stellar flux and $R_\ast$ the stellar radius.

\begin{figure}[t]
      \centering
      \includegraphics[width=.61\linewidth]{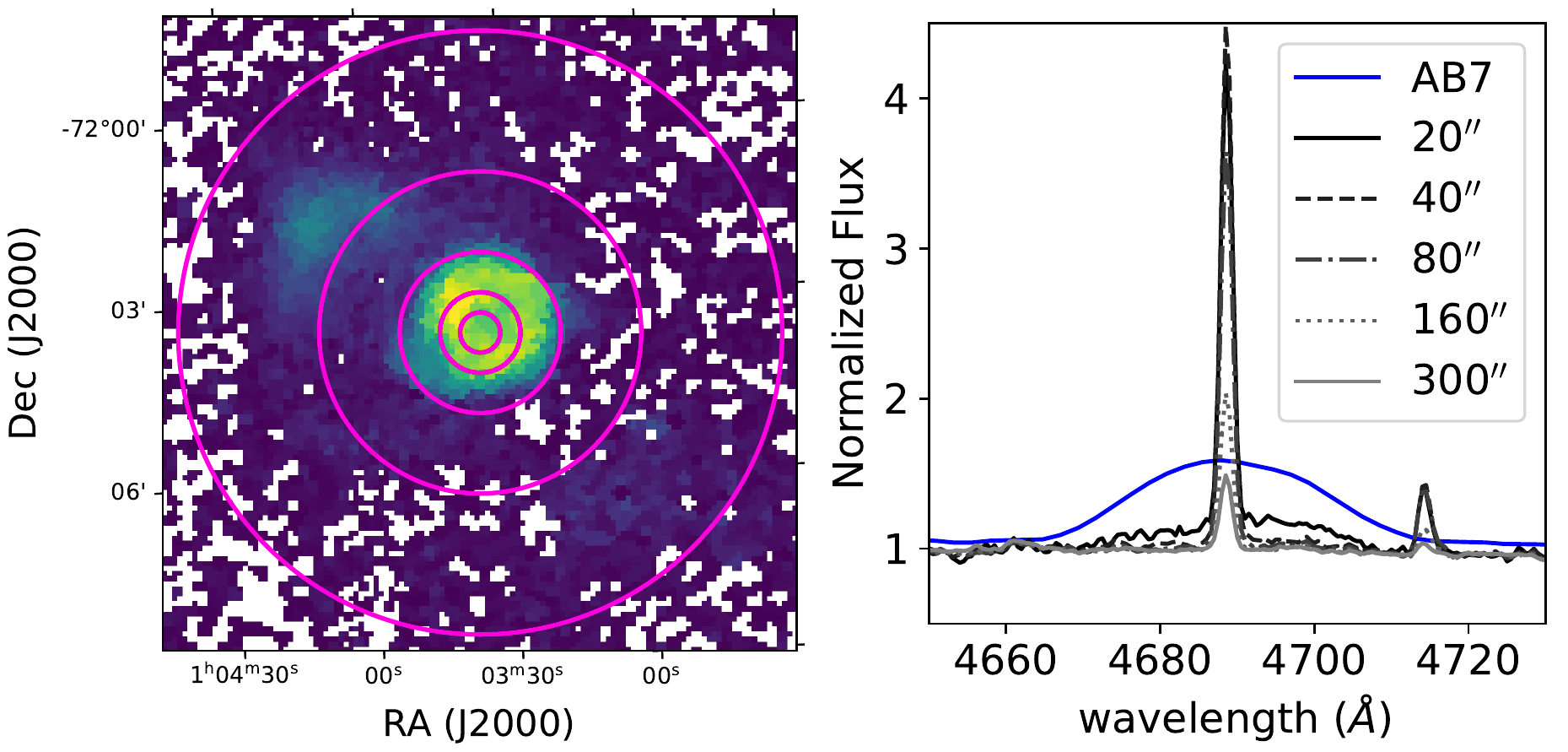}
      \hfill
      \includegraphics[width=.37\linewidth]{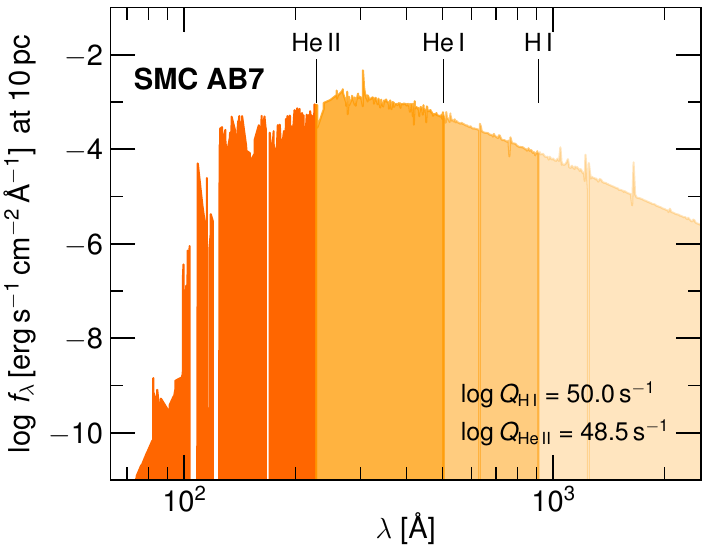}
      \caption{\textit{Left:} AB7 fiber apertures studied from the LVM. \textit{Center:} Normalized 4686 $\AA$ flux of the star and the different fiber apertures of the LVM. \textit{Right:} Intrinsic flux of the WN3h component in the AB\,7 binary.
      }
      \label{fig:ab7}
\end{figure}
Thanks to SDSS V \citep{Kollmeier2025}, the availability of the LVM wide field IFU data now enables us to cross-check the spectra from the individual WN3 targets as well as their inferred quantities with nebular diagnostics. LVM covers a field of view of 0.5 degrees with 1801 hexagonally packed fibers of 35.3’’ apertures, with a spectral coverage 3600-9800 $\AA$, and a spectral resolution of R$\sim$4000. 
We have studied the integrated spectra from the fiber apertures around the SMC WN3 stars with angular radii of 20’’, 40’’, 80’’, 160’’, and 300’’, corresponding to a diameter of 12, 24, 48, 96 and 181 pc (see left panel in Figure\,\ref{fig:ab7}). 
From the nebular emission, we estimated Q$_\mathrm{He_{II}}$ using the formula from \citet{Kehrig2015}:
\begin{equation}\label{eq:flux}
    \text{Q}(\text{HeII})=L_{\text{HeII}}/[j(\lambda 4686)/\alpha_{\text{B}}(\text{HeII})]\approx L_{\text{HeII}}/3.66\times10^{-13} ,
\end{equation}
where $L_{\text{HeII}}$ is the line luminosity obtained from the LVM spectra, $j(\lambda 4686)$ is the photon energy at 4686$\,\AA\,$, and $\alpha_{\text{B}}(\text{HeII})$ the recombination rate coefficient assuming case B recombination and $T_\text{e}\sim2\times10^{4}$K \citep{Osterbrock2006}.

\section{Results and conclusions}

\noindent%
In all cases, the stellar emission is strongly diluted in the LVM data. An example is shown in the right panel of Figure\,\ref{fig:ab7}, were we plot the normalized spectra around He\,II\ $4686\,\AA$ for AB7 and the corresponding LVM fiber apertures. While still present in the smallest 20'' aperture, the stellar component completely vanishes at 40'', corresponding to a diameter of 24 pc. In \citet{Gonzalez-Tora2025} we show that the stellar contribution for the rest of the targets is always completely diluted at an aperture of only 24 pc. These results imply that WN3 stars at higher distances in unresolved environments can be hard or even impossible to detect. 

\begin{figure}[t]
      \centering
      \includegraphics[width=.58\linewidth]{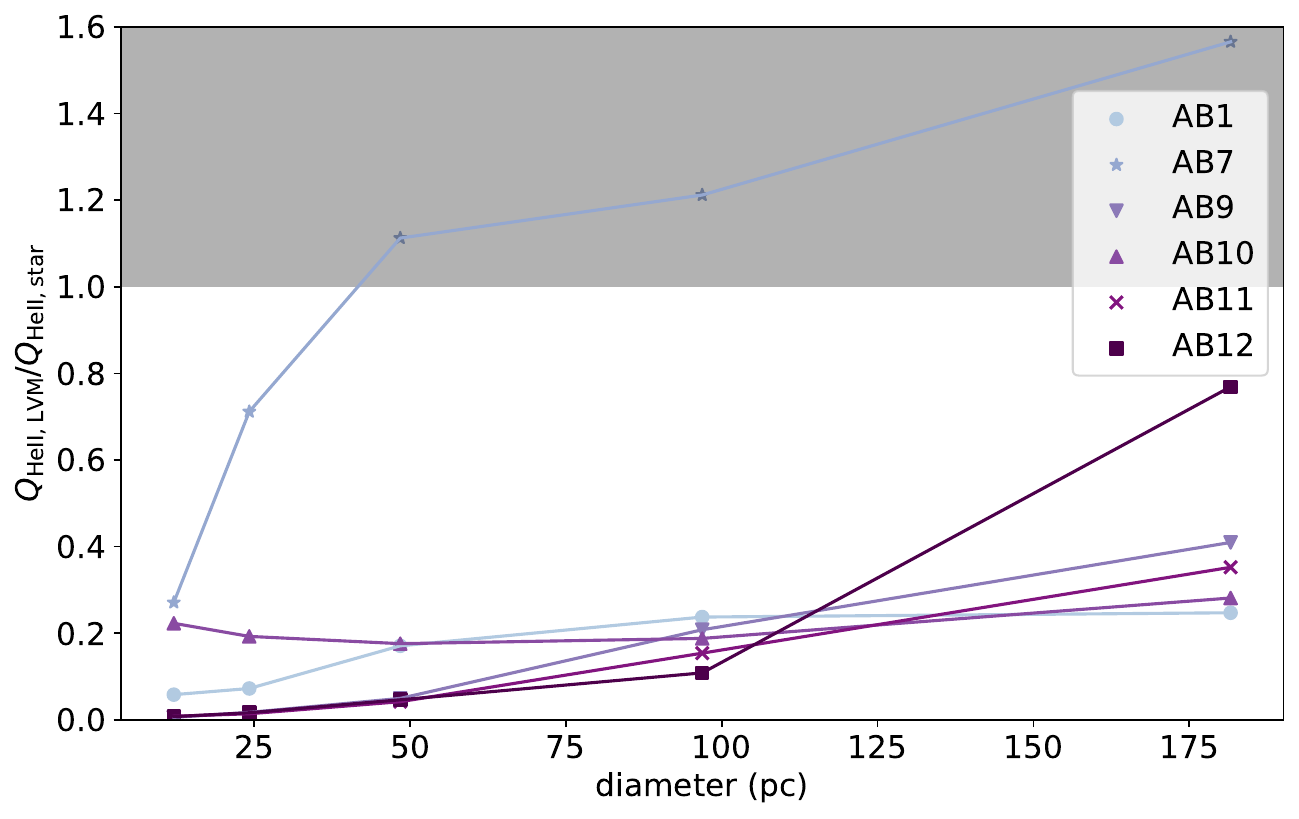}
      \caption{Ratio of Q$_{\mathrm{HeII}}$ for the different stars and the corresponding LVM integrated region with respect to the diameter of the LVM aperture. The Q$_{\mathrm{HeII, star}}$ ratio is obtained by the PoWR models, while the Q$_{\mathrm{HeII, LVM}}$ is calculated using Eq.~\ref{eq:flux}, using only for the narrow nebular component at each aperture. 
      }
      \label{fig:qheii}
\end{figure}
However, their non-detection does by no means imply that such stars could not contribute significant numbers of ionizing photons. Using again the LVM data, we can compare how much of the intrinsic Q$_\mathrm{He_{II}}$ is visible in the observed nebular He II emission lines. 
Figure\,\ref{fig:qheii} shows the ratio of ionizing fluxes Q$_\mathrm{He_{II},LVM}/\mathrm{Q}_\mathrm{He_{II},star}$ with respect to the diameter of the LVM integrated fiber in pc. We see that for all cases -- except AB\,7 with d$>$48 pc -- the star alone produces more than enough photons to produce the observed emission in largest aperture (i.e., Q$_\mathrm{He_{II},LVM}/\mathrm{Q}_\mathrm{He_{II},star}<1$). For AB7 at d$>$48 pc, there is a blending with other nearby sources of He\,II ionizing flux, possibly from within NGC\,395.

Our results demonstrate that WN3 stars with comparably thin winds and thus weak emission lines are easy to hide, despite producing enough photons to reproduce the observed nebular He\,II emission. Current evolution models cannot reproduce these observed WR stage, in particular at SMC metallicities and below, thereby causing a major deficiency in current population synthesis. Solving the origin puzzle of the WN3 stars and other related objects such as WN2 stars \citep{Sander2025} is urgently needed and will likely have major impacts on stellar population predictions and cosmological simulations of galaxy evolution.
\bibliographystyle{apj}
\bibliography{Sample}

\end{document}